\documentclass[fleqn,usenatbib]{mnras}

\usepackage{newtxtext,newtxmath}
\usepackage{bm}
\usepackage{amsmath}

\usepackage[T1]{fontenc}

\DeclareRobustCommand{\VAN}[3]{#2}
\let\VANthebibliography\thebibliography
\def\thebibliography{\DeclareRobustCommand{\VAN}[3]{##3}\VANthebibliography}

\usepackage{graphicx}	
\usepackage{amsmath}	

\title[Mass-Ratio Distribution I]{The Frequency and Mass-Ratio Distribution of Binaries in Clusters I: Description of the method and application to M67. }

\author[M. D. Albrow \& I. H. Ulusele]{Michael D. Albrow\thanks{E-mail:Michael.Albrow@canterbury.ac.nz},
Isaac H. Ulusele\\
\\
$^{1}$School of Physical and Chemical Sciences, University of Canterbury, Private Bag 4800, Christchurch, New Zealand\\
}

\date{Accepted XXX. Received YYY; in original form ZZZ}

\pubyear{2022}

\begin{document}
\label{firstpage}
\pagerange{\pageref{firstpage}--\pageref{lastpage}}
\maketitle

\begin{abstract}
We present a new method for probabilistic generative modelling of stellar colour-magnitude diagrams (CMDs) 
to infer the frequency of binary stars and their mass-ratio distribution. 
The method invokes a mixture model to account for overlapping populations of single stars, binaries and outliers
in the CMD. We apply the model to Gaia observations of the old open cluster, M67, and find a frequency
$f_B(q > 0.5) = 0.258 \pm 0.019$ for binary stars with mass ratio greater than 0.5. The form of the
mass-ratio distribution function rises towards higher mass ratios for $q > 0.3$.

\end{abstract}




\section{Introduction}

%
%
%

Studies have shown that perhaps 40 - 90 percent of field stars in the Milky Way are born in clusters or associations \citep{Bressert2010,Ward2020}. 
Binary stars play an important role in the evolution of such systems. Through 3-body or 4-body  gravitational interactions, binaries can enhance the dispersal of a star cluster through ejection of stars.
In  dense clusters, binaries provide an energy reservoir that prevents runaway core collapse through the conversion of binding energy to kinetic energy in three-body encounters. Binary stars in a cluster shape its structure and evolution, which in turn shapes the structure and evolution of the binary systems \citep{Hut1992,Hurley2005}.

Counts of binary stars in clusters have been done by various authors using either radial velocities, photometric colour-magnitude diagrams, or photometric times-series analysis.

CMD photometric methods generally apply some criterion to separate single from binary stars, count the binary stars in particular CMD locations (sometimes comparing this to models), and apply some mass-ratio function to compute the total binary frequency. For some examples of this general approach see \citet{Milone2012}, \citet{Ji2013} for globular clusters, \citet{Elson1998} and \citet{Li2013} for young massive LMC clusters, and \citet{Sollima2010} for galactic open clusters.

Radial velocities can be used to detect binary stars through their orbital motion, and can be used as a cluster 
membership criterion. The detected radial velocity binary population must be corrected statistically for declining detection efficiency for low orbital inclination to derive the cluster binary frequency. 
Radial velocity studies, e.g. the WIYN Open Cluster Study \citep{Mathieu2000}, have generally been limited to stars near the top of the main sequence or brighter.

Time series analysis to detect variable stars, including binaries, has been used by \citet{Albrow2001} for the globular cluster 47 Tucanae. Inference of a binary frequency is, however, fairly indirect, relying on assumptions such as the contact binary lifetime.

The mass ratio distribution of field binary stars has been determined from radial velocity surveys by several authors.
\citet{Duquennoy1991} found a distribution that peaks around $q = 0.25$ and declines towards higher mass ratios.
In contrast, \citet{Fisher2005} infer a flat distribution with a sharp upturn towards higher mass ratios. 
For globular clusters, \citet{Milone2012} find flat or gently rising distributions with $q$ (above $q = 0.5$). 
The model that we will construct below allows for these possibilities.
As noted by \citet{Milone2012}, all the measured mass ratio distributions for field stars and clusters are fundamentally at odds with
random draws of secondaries from the initial mass function.

Overall, it has been found that the frequency of binary stars in open clusters can range from  25 - 70 \%. For globular clusters, the binary frequency is much lower, certainly less than 25 \%, and possibly less than 5 \%. Prior to this paper, all estimates rely on assumptions about the form of the mass ratio ($q$) distribution at low $q$. Estimates are also complicated further by the well-established phenomenon that binaries are more centrally concentrated than single stars. 

M67 (NGC 2682) is a rich old Milky Way open cluster located at
coordinates
RA = 08:51:23, DEC = +11:49:02 (J2000) and a distance of 860 pc.
It has been extensively studied photometrically (e.g \citet{Yadav2008,Sarajedini2009,Gao2018ApJ}), and has been the target or several spectroscopic surveys \citep{Pasquini2012,Geller2021}.
The binary star frequency has been estimated as >38\% \citep{Montgomery1993}
>26\% \citep{Gao2018ApJ}, >45\% \citep{Davenport2010}
$\sim$ 50\% \citep{Fan1996}, $34 \pm 3$\% \citep{Geller2021}.

The layout of the remainder of this paper is as follows. In Section~\ref{sec:Gaia} we discus Gaia observations of M67, 
and our filtering to achieve a clean CMD. In Section~\ref{sec:model} we develop a parameterised generative model for the 
CMD, which we solve in Section~\ref{sec:results}.

\subsection{M67}

\section{Gaia photometry}\label{sec:Gaia}

\subsection{Selection}\label{sec:Selection}

Standard Gaia EDR3 photometry is available in three photometric passbands, 
$G_{BP}$ and $G_{RP}$ integrated from low resolution spectra, and $G$ measured from the 
astrometric field CCDs \citep{Evans2018, Riello2021}. Hereafter we refer to these
as $B$, $R$ and $G$.
We first made the flux-excess corrections to the $B$ and $R$ magnitudes as recommended
by \citet{Riello2021}. For our data, these corrections were almost negligible.
There is a known problem with overestimated  Gaia $B$ flux for  
faint red stars (see Section 8.1 in \citet{Riello2021}) so for our
analysis we use colour-magnitude diagrams in $(G-R,G)$.

\begin{figure}
	\includegraphics{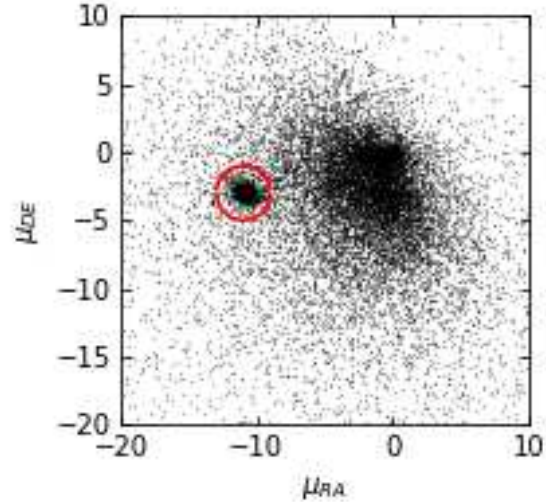}
    \caption{Proper motion of stars in a 1 deg radius cone centred on M67. The red circle encloses our initial selection of cluster stars.}
    \label{fig:pm1}
\end{figure}

Initially we selected all Gaia EDR3 data from a cone of radius 1 deg centered on M67 (NGC 2682). This data included the cluster as well as many foreground and background stars. We next made a coarse cut on proper motion, selecting those stars that had a proper motion $(\mu_\alpha, \mu_\delta)$ that lay within a circle of radius 2 mas yr$^{-1}$ centred on (-11, -3)  mas yr$^{-1}$ (Fig.~\ref{fig:pm1}). We next filtered out stars with a parallax uncertainty greater than 20$\%$ and fitted a 5-D gaussian 
in galactic latitude ($l$), galactic longitude ($b$), $\mu_\alpha$, $\mu_\delta$, and parallax ($\varpi$) to the remainder. By trial and error, we imposed a cut on this 5-D distribution at a level for which we were satisfied with a moderately clean colour-magnitude diagram, Fig~\ref{fig:CMD1}. The parameter boundaries for the adopted cuts are listed in Table~\ref{table:data_cuts}. This process has resulted in a very clean CMD, with less contamination than that of \citet{Yadav2008}, albeit not as deep as that study. It appears similar in quality to the M67 Gaia CMD derived by \citet{Gao2018ApJ} using machine-learning methods. 

The final CMD shows a main sequence with an obvious population of binary stars distributed up to 0.75 mag above the main-sequence ridge line, plus a few outliers that may be triple systems, or evolved binaries. Binary systems with two equal-mass main-sequence stars appear 0.75 mag above the main sequence. Binaries with lower mass ratios appear redder and fainter than the equal-mass case, see for instance the tracks in Fig.~3 of \citet{Elson1998}.

There is some evidence for mass segregation in M67. \citet{Fan1996} found a half-mass radius of 7 arcmin for blue stragglers, 10 arcmin for upper main sequence stars, and 12 arcmin for the lower main sequence. 
This is confirmed from radial velocity studies \citep{Mathieu1986,Geller2021}. We might therefore expect main-sequence binaries to be more centrally concentrated than single stars.
Our analysis covers stars drawn from a radius of $\sim 30$ arcmin, so effectively averages over any mass segregation
of binaries. The tidal radius of M67 is estimated to be 42 arcmin \citep{Fan1996,Francic1989}, and its core radius is
8.24 arcmin \citep{Davenport2010}.

\begin{table} 
	\centering
	\caption{5-dimensional selection cuts applied to the Gaia data.} \label{table:data_cuts}
	\begin{tabular}{lcc}
	\hline
	& Minimum & Maximum \\
	\hline
	$l$ & 215.18 & 216.20 \\
	$b$ & 31.48 & 32.37 \\
	$\varpi$ (mas) & 1.006 & 1.314 \\
	$\mu_\alpha$ (mas/yr) & -11.77 & -10.06 \\
	$\mu_\delta$ (mas/yr) & -3.98 & -2.04 \\
	\hline
	\end{tabular}
\end{table}

For the analysis presented below, we restricted ourselves to considering stars that are clearly on the main sequence
or are binaries consisting of two main-sequence stars. We have manually removed the few stars that are clearly above the
equal-mass-binary main-sequence.
Additionally we have retained only stars where the primary (or only) component lies within $13.5 < G < 18$.
These cuts are shown as magenta curves in Fig.~\ref{fig:CMD1}, and were found necessary in order that our binary star distribution
can be modelled below with the same primary-star mass function as the single stars.

\subsection{Covariant uncertainties}

Each 2-dimensional data point on the CMD, $(G-R,G)$ is formed from G and R
measurements with independent uncertainties, but the $G-R$ combination is not independent of $G$.
Consequently a $(G-R,G)$ datum has an associated covariance matrix
\begin{equation}
\mathbfss{S} = 
\begin{bmatrix}
\sigma_{G-R}^2 & \sigma_{G-R,G}  \\
\sigma_{G-R,G}  & \sigma_G^2
\end{bmatrix},
\end{equation}
where $\sigma_{G-R}^2 = \sigma_G^2 + \sigma_R^2$ and it can be shown (see Appendix~\ref{appendixA}) that the off-diagonal elements, $\sigma_{G-R,G} = \sigma_G^2$.
As viewed in the $(G-R,G)$ plane, each data point can be considered as having a
tilted bivariate gaussian probability distribution.

\subsection{Isochrone fitting}

\begin{figure}
	\includegraphics{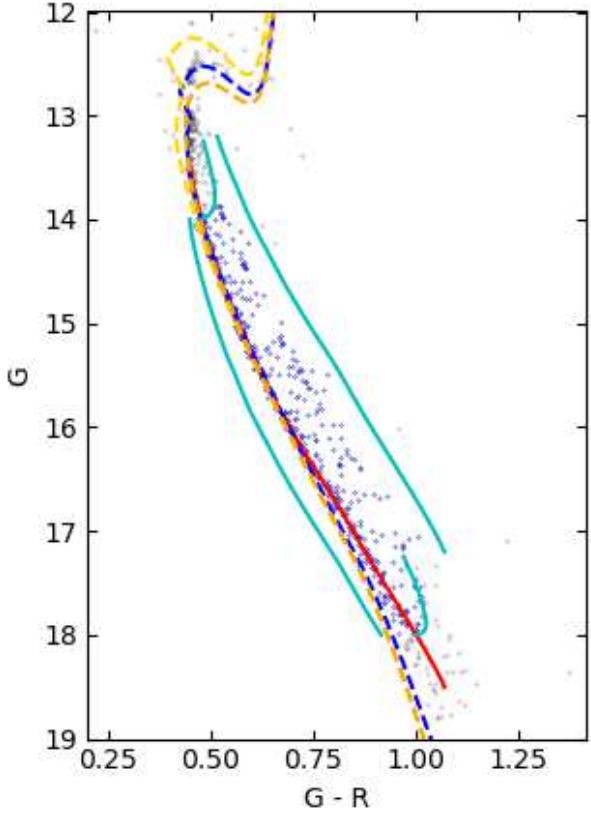}
    \caption{Colour-magnitude diagram for stars passing the filters as described in Section~\ref{sec:Selection}. The dashed blue line is the MESA isochrone for an age of
    5 Gyr, The red line shows the colour-corrected isochrone. 
    Gold and orange dashed lines show isochrones for 4 Gyr and 6 Gyr respectively.
    Cyan curves show the lower and upper cuts applied
    so that binary-star primary masses are drawn from the same population as single stars, and some broader cuts to reject obvious outliers.
    }
    \label{fig:CMD1}
\end{figure}

To model the stellar population of the cluster, we rely on having an isochrone that accurately describes the main sequence magnitude and colour as a function of mass. We used isochrones
from the MESA Isochrones and Stellar Tracks project\footnote{http://waps.cfa.harvard.edu/MIST/}
\citep{Dotter2016,Choi2016,Paxton2011,Paxton2013,Paxton2015}. After some trial and error, we adopted a 5 Gyr isochrone with $[{\rm Fe/H}] = +0.06$ as being the best fit to the upper main sequence and 
turnoff.  This is consistent with the metallicty determinations of \citet{Hobbs1991} ($[{\rm Fe/H}] = -0.04 \pm 0.12$) and \cite{Onehag2014A&A} ($[{\rm Fe/H}] = 0.06$). The age, however,
is long compared with most recent determinations, which are generally close to 4 Gyr.
The age inferred for 
M67, however, depends on the adopted theoretical isochrone calculations (see Fig. 10 of \citet{Yadav2008}), in particular the treatment of convective overshooting.
In Fig.~\ref{fig:CMD1} we shown the M67 Gaia CMD with MESA isochrones for 4, 5 and 6 Gyr
and $[{\rm Fe/H}] = +0.06$. We based our fit on the main-sequence turnoff region,
and found that only the 5 Gyr isochrone provided a  correct width for the subgiant branch and matched the "hook" at the top of the main sequence. 
The specific choice of isochrone is inconsequential for the purpose of this paper since we will eventually marginalise over all parameters associated with the mass distribution.

The MESA isochrones deviate from the observed main-sequence ridge line
near the bottom of the main sequence. We introduce a colour correction to the isochrone
below $G = 16.5$, implicitly adopting the $G$ vs mass relation from the isochrone.

\subsection{Limits on binary star detection}

\begin{figure}
	\includegraphics[width=\columnwidth]{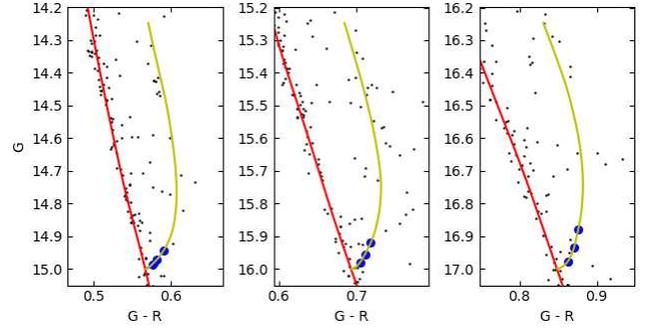}
    \caption{Cutout sections of the final colour-magnitude diagram near different G magnitudes. The red lines are the adopted isochrone with colour correction applied. The yellow lines are tracks for binary stars with the same primary mass and varying mass ratio.
    The blue dots mark the locations $q$ = (0.4, 0.5, 0.6).
    }
    \label{fig:CMD_q_iso}
\end{figure}

Depending on the quality of the photometry and the chosen passbands, a given CMD will have a threshold in $q$ below which binaries
are practically indistinguishable from single stars.
In Fig.~\ref{fig:CMD_q_iso} we show sections of our final trimmed CMD data, along with the colour-corrected isochrone for the main sequence and tracks representing
binary stars with a fixed primary mass and different values of $q$.
We indicate the locations of binary stars with $q = $ (0.4, 0.5, 0.6).
For our data, $q = 0.5$ is a sensible threshold to adopt, as we
perceive that such binary stars 
are sufficiently displaced from the main sequence to be recognised as such.
The model we will develop in the following section does not depend on this threshold, but it is important for the interpretation of 
the model outputs.

\section{Model}\label{sec:model}

In this section we develop a probabilistic generative model for the main-sequence region of a colour-magnitude diagram.
Our aim is to develop a probability for any location on the CMD as a function of a set of numerical parameters, and then to sample that probability distribution for the observed data in order to learn the probability distributions for the parameters. Our method is strongly influenced by the mixture modelling approach of \citet{Hogg2010}.
We refer the reader to \citet{Taylor2015} for a comprehensive example of constructing a generative probablistic model, in that case for the bimodal colour-colour distribution for field galaxies from the Galaxy And Mass Assembly (GAMA) survey.

Our Gaia data is the set ${\bm D} = ({\bm D_k})$ for stars $k$, with each  ${\bm D_k} = (G_k-R_k,G_k)^T$.  
Assuming that our model for the data can be described by a vector of parameters ${\bm \theta}$ (to be described), then
from Bayes theorem the probability distribution for ${\bm \theta}$,

\begin{equation}
    P({\bm \theta} | {\bm D}) = \frac{P({\bm D} | {\bm \theta})  P({\bm \theta})}{Z},
\end{equation}

where $P({\bm D} | {\bm \theta})$ is the likelihood function, $P({\bm \theta})$ is the prior for ${\bm \theta}$ and $Z$ is the marginal likelihood, a constant for a given ${\bm D}$ and model.

We recognize that the majority of CMD data points are either single stars or binaries, but that there may be some "bad" data
points that would not be represented by a combination of single and binary star distributions in the CMD.
We adopt a generative mixture model \citep{Hogg2010}, where each star has a probability $f_{\rm B}$ of being a binary, $f_{\rm O}$ of being an outlier, 
and $(1- f_{\rm B} - f_{\rm O})$ of being a single star.
This approach is a natural treatment for the overlapping populations close to the
main-sequence ridge line, and explicitly does not require the labelling of an individual
star as being a member of a specific population. By design, our data sample contains few outliers,
but we retain the $f_O$ term for generality of the model description. 

The likelihood function,
\begin{equation}
P({\bm D_k} | {\bm \theta}) = (1 - f_{\rm B} - f_{\rm O}) P_{\rm S}({\bm D_k} | {\bm \theta}) + f_{\rm B} P_{\rm B}({\bm D_k} | {\bm \theta}) + 
f_{\rm O} P_{\rm O}({\bm D_k} | {\bm \theta}),
\end{equation}
where $P_{\rm S}({\bm D} | {\bm \theta}), P_{\rm B}({\bm D} | {\bm \theta}), P_{\rm O}({\bm D} | {\bm \theta})$ are respectively the
likelihood functions for single stars, binaries and outliers. We describe each of these in turn below.
The total likelihood is the product of the individual star likelihoods so that
\begin{equation} \label{eqn:prob}
\ln P({\bm D} | {\bm \theta}) = \sum_k \ln P({\bm D_k} | {\bm \theta}) .
\end{equation}

\subsection{Error-bar rescaling}

It is common in observational studies for measurement uncertainties to be underestimated (or overestimated). In practice, this is equivalent to assuming that the underlying true distribution
has an intrinsic width. 
Such a width could arise from effects such as differential reddening or
small systematic errors in the photometry.
We allow for the possibility of an intrinsic width or incorrect
error bars in our model by adopting a free parameter, $h$, that 
scales all data error bars. In general this manifests as 
$\mathbfss{S}_{G,G-R} \rightarrow h^2 \mathbfss{S}_{G,G-R}$ in the probability calculations
that follow. From examination of Fig.~\ref{fig:CMD1}, it appears that the data may become more noisy 
towards the lower main sequence, so we allow $h$ to vary linearly along the main sequence as
$h = h_0 + h_1 (G - G_0)$, where we adopt $G_0 = 16.0$. Error bar scaling thus introduces two parameters to our model.

\subsection{Outliers}

We begin with the distribution of outliers, since that is the simplest of the three individual likelihoods. In what follows, we adopt the notation
\begin{equation}
\mathcal{N}({\bm x} ,\mathbfss{S}) \equiv
\frac{1}{2 \pi \sqrt{\det (\mathbfss{S})}} 
\exp\left(  -\frac{1}{2} {\bm x}^T 
\mathbfss{S}^{-1} {\bm x} \right)
\end{equation}
for a normalised bivariate gaussian centred at $(0,0)$  with covariance matrix $\mathbfss{S}$,  evaluated at $x$.
We also note that a probability density in $x$ should strictly be denoted by the
derivative, $dP/dx$. For brevity, we adopt the looser notation, $P(x)$. 

From the standard rules of probability we can expand the outlier likelihood function
\begin{equation}
P_{\rm O}({\bm D_k} | {\bm \theta})  = 
\int P({\bm D_k} | {\bm D'}) P_{\rm O}({\bm D'} | {\bm \theta}) \, {\rm d}^2{\bm D'}.
\end{equation}
The first term in the integral is the (gaussian) probability of 
the given data point ${\bm D_k}$ given a "true" CMD location ${\bm D'}$. The second term is the
probability distribution on the CMD for outliers, for which we 
adopt a very broad bivariate gaussian. Consequently, the integral becomes a
convolution
\begin{align}
P_{\rm O}({\bm D_k} | {\bm \theta}) & = 
\int \mathcal{N}({\bm D_k} - {\bm D'},h^2 \mathbfss{S}_k) 
\mathcal{N}({\bm D'} - {\bm D_O},{\bm S_0}) \, {\rm d}^2{\bm D'} \\
& = \mathcal{N}({\bm D'} - {\bm D_k},h^2 \mathbfss{S}_k) \ast
\mathcal{N}({\bm D'} - {\bm D_O},{\bm S_0}) \\
& = 
\mathcal{N}({\bm D_k} - {\bm D_O},h^2 \mathbfss{S}_k+{\bm S_O}),
\end{align}
and the likelihood is thus a bivariate gaussian.
In principle, ${\bm D_O}$ and  ${\bm S_O}$ could be regarded as elements
of ${\bm \theta}$, but in practice it is sufficient to set them as constants. Here we adopt
${\bm D_O} = (0.75,16)$ and ${\bm S_O} = {\rm diag}((0.75)^2,(4.0)^2)$.
Generally $\mathbfss{S}_k$ is negligible compared to ${\bm S_O}$ and can
be removed from the final covariance with no effect. 

The above description is certainly an imprecise representation of the outlier distribution but,
as noted by \citet{Hogg2010}, much of the power of a mixture model such as this one
comes from simply including an outlier distribution, even if it is inaccurate
in detail.

\subsection{Single stars}

We assume that the true locations of single stars in the CMD lie along the isochrone line,
and are drawn from some mass distribution. This mass distribution is truncated at the
top of the main sequence, and fades away more gradually at the bottom of the main sequence
due to declining Gaia detection efficiency with increasing G magnitude. We model this 
observational mass distribution as

\begin{equation} \label{eqn:single_mass_function}
\begin{split}
P(M|\gamma,k,M_{\rm 0},M_{\rm max}) = {} & 
C_M M^{-\gamma} \times \frac{1}{1 + e^{-k (M-M_0)} } \\
& \times H(M_{\rm max}-M).
\end{split}
\end{equation}

Here, the first term is a power law in $M$. The second
term is a logistic function that
acts as a smoothed step, centred at $M_0$ and with characteristic width $1/k$, to control
the bottom of the mass distribution.  The top of the distribution is
subject to a sharp cutoff by the the third term, a Heaviside step function at mass $M_{\rm max} = 1.186 \, {\rm M}_\odot$, corresponding to G=13.5.
The normalisation parameter $C_M$ is computed numerically so that $\int_{-\infty}^\infty P(M|\gamma,k,M_{\rm 0},M_{\rm max}) dM = 1$.
For our M67 data, we finally opted to apply a hard cut near the bottom of the main sequence, so could have replaced the logistic function with a second step function, however we retain this form for generality of the model.

Since there is a one-to-one mapping, $D'(M)$, from a given $M$ to a location on the CMD
the single-star likelihood function can be expanded as
\begin{align}
P_{\rm S}({\bm D_k} | {\bm \theta}) & = \int P({\bm D_k} | {\bm D'}) P({\bm D'}|M)  P(M | {\bm \theta}) \, {\rm d}^2{\bm D'} \,dM \\
& = \int P({\bm D_k} | {\bm D'(M)}) P(M | {\bm \theta}) \, dM \\
& = \int \mathcal{N}({\bm D_k} - {\bm D'(M)},h^2 \mathbfss{S}_k) P(M | {\bm \theta}) \, dM
\end{align}
We can perform this integration by breaking the mass distribution function into a sum of discrete 
mass steps of width $\Delta M$ and height $P(M_l | {\bm \theta})$ located at masses $M_i$,
\begin{align}
P_{\rm S}({\bm D_k} | {\bm \theta}) & = \sum_l \mathcal{N}({\bm D_k} - {\bm D'(M_l)},h^2 \mathbfss{S}_k)  
P(M_l | {\bm \theta}) \Delta M .
\end{align}

In practice, results from using this expression are equivalent to employing the expression we will develop in the following section for binary stars, with a mass ratio set to be close to zero.

\subsection{Binary stars}

We assume that the more massive star, $M_1$, in each binary system is drawn from the same distribution as the single-star mass function, Eqn~\ref{eqn:single_mass_function}. Each secondary star has a mass $M_2 = q M_1$, where $q$ is drawn from a parameterised probability density. After testing various forms for this distribution, we adopted a general cubic polynomial,
\begin{equation} \label{eqn:q_function}
P(q|a_k,\dot{a}_k,M; k = 1..3) = \tilde{P}_0(q) + \Sigma_{k=1}^{3} (a_k + \dot{a}_k (M-M_{\rm ref})) \tilde{P}_k(q),
\end{equation}
where $\tilde{P}_k(q)$ are the shifted Legendre polynomials, an orthogonal basis set over the range $0 < q< 1$. This general form allows the distribution to bend up or down at either or both ends. The form of the distribution as written requires no 
normalisation since
\begin{equation}
\int_0^1 \tilde{P}_k(x) dx = 
\begin{cases} 
1, & k = 0 \\
0, & k > 0.
\end{cases}
\end{equation}
By including the derivative parameters $\dot{a}_k \equiv \partial a / \partial M$, the shape of the distribution is allowed 
to vary with mass, around some reference value $M_{\rm ref}$, which we choose as the centre of the mass range covered by the CMD main sequence. 

We consider this general model for the distribution as described, and also the more restricted versions with $\dot{a}_k$ set to zero.

To compute the likelihood, we represent the probability densities for $M_1$ and $q$ as linear combinations of 50
predefined gaussian basis functions,
\begin{align}
& P(M_1)  = \sum_i a_i \mathcal{N}(M_1 - M_{1,i},\sigma_{M}) \\
& P(q)  = \sum_j b_j \mathcal{N}(q - q_j,\sigma_{q})
\end{align}
where the centroids, $M_{1,i}$ and $q_j$ are equally spaced across the mass and mass-fraction ranges, $0.1 <= M_1/M_\odot < = 1.1$ and  $0 < q <= 1$. The gaussian widths are set to 
$\sigma_{M_1} = 0.01$ and $\sigma_q = 0.01$, which we found by trial-and-error to produce a 
good representation of the underlying functions. The choice of gaussians for these basis functions will become apparent below.

It is a requirement of the formalism that follows that all $a_i$ and $b_j$ are zero or positive.
The coefficients $a_i$ and $b_j$ can be computed analytically as linear fits to numerical representations of Eqns~\ref{eqn:single_mass_function} and \ref{eqn:q_function}.
However, the optimal linear fit coefficients are not guaranteed to be positive. 
Instead we use the non-negative least squares algorithm from \citet{Lawson1974},
which iterates to a solution.

A model magnitude for a binary star is obtained by adding the  fluxes
of the two components in the appropriate bandpass. This is done by interpolating
magnitudes for masses $M_1$ and $q M_1$ from the isochrones, converting magnitudes
to fluxes, adding the fluxes, and reconverting to magnitudes. This procedure is performed separately for 
$G$ and $R$, then the resulting values are combined to form $(G-R, G)$.

Each combination of $(M_i,q_j)$ forms a normalised bivariate gaussian in $(M,q)$ space with
covariance matrix, $\mathbfss{S}_{M,q} = {\rm diag}(\sigma_M^2,\sigma_q^2)$. This transforms to a 
covariance matrix in $(G-R,G)$ space, $\mathbfss{S}_{G-R,G} = \mathbfss{J} \mathbfss{S}_{M,q} \mathbfss{J}^{T}$, 
where the Jacobian
\begin{equation}
\mathbfss{J} = \begin{bmatrix}
\frac{\partial G-R}{\partial M} & \frac{\partial G-R}{\partial q} \\
\\
\frac{\partial G}{\partial M} & \frac{\partial G}{\partial q}
\end{bmatrix}.
\end{equation}
The partial derivatives in the Jacobian are not dependent on the data or model parameters and can be pre-computed for any given $(M,q)$ by interpolating
colours and magnitudes from the isochrone.
The combination of all the $(M,q)$ bivariate gaussians map to a set of tilted overlapping bivariate gaussians in $(G-R,G)$ space. 

Similar to what we have done previously, we expand the
binary-star likelihood
\begin{equation}
P_{\rm B}({\bm D_k} | {\bm \theta})  = 
\int P({\bm D_k} | {\bm D'}) P_{\rm B}({\bm D'} | {\bm \theta}) \, {\rm d}^2{\bm D'}.
\end{equation}
This can be further decomposed as
\begin{align}
P_{\rm B}({\bm D_k} | {\bm \theta})  & = 
\sum_{i,j} a_i b_j \int P({\bm D_k} | {\bm D'}) P_{\rm B}({\bm D'} | M_i, q_j) \, {\rm d}^2{\bm D'} \\
& = \sum_{i,j} a_i b_j \int \mathcal{N}({\bm D_k} - {\bm D'},h^2 \mathbfss{S}_k) 
\mathcal{N}({\bm D'} - {\bm D_{ij}},\mathbfss{S}_{ij}) \, {\rm d}^2{\bm D'} 
\end{align}
where ${\bm D_{ij}} \equiv {\bm D}(M_i,q_j)$ and $\mathbfss{S}_{ij} \equiv \mathbfss{J} \mathbfss{S}_{M_i,q_j} \mathbfss{J}^{T}$.
Here, the choice of gaussian bases for $M$ and $q$ becomes apparent. 
The integral is again a convolution that can be evaluated analytically, so that
\begin{equation}
P_{\rm B}({\bm D_k} | {\bm \theta}) = \sum_{i,j} a_i b_j 
\mathcal{N}({\bm D_k} - {\bm D_{ij}},h^2 \mathbfss{S}_k + \mathbfss{S}_{ij}). 
\end{equation}

\subsection{Total likelihood} \label{sec:total_likelihood}
Expanding Eqn.~\ref{eqn:prob}, 
\begin{equation}
\begin{split}
\ln P({\bm D} | {\bm \theta}) = {} 
& \sum_k \ln \left\{ (1 - f_{\rm B} - f_{\rm O}) P_{\rm S}({\bm D_k} | {\bm \theta}) + f_{\rm B} P_{\rm B}({\bm D_k} | {\bm \theta}) \right. \\ 
& + \left. f_{\rm O} P_{\rm O}({\bm D_k} | {\bm \theta}) \right\}.
\end{split}
\end{equation}
Inserting the individual likelihoods developed above, we can write this in a compact nested form,
\begin{equation} \label{eqn:final_ln_prob}
\ln P({\bm D} | {\bm \theta}) = \sum_k {\rm LSE} \left\{ A_k, B_k, C_k \right\},
\end{equation}
where LSE is the log sum exp function, 
\begin{equation}
    {\rm LSE}(a,b,...) \equiv \ln \left( e^a + e^b  + ...\right),
\end{equation}
and 
\begin{align}
A_k = {} &  -\frac{1}{2} ({\bm D_k} - {\bm D_O})^T \mathbfss{S}_0^{-1} ({\bm D_k} - {\bm D_O})
    + \ln \frac{f_{\rm O}}{2 \pi \sqrt{\det (\mathbfss{S}_O)}} ,\\
\begin{split}
B_k = {} & \underset{l}{\rm LSE} \left\{ - \frac{1}{2} ({\bm D_k} - {\bm D}(M_l))^T (h^2 \mathbfss{S}_k)^{-1} ({\bm D_k} - {\bm D}(M_l)) \right. \\
 & \left. +  \ln \frac{(1 - f_B - f_{\rm O}) P(M_l|{\bm \theta}) \Delta M }{2 \pi \sqrt{\det ( h^2 \mathbfss{S}_k) } } \right\}, 
\end{split} \\
\begin{split}
C_k = {} & \underset{ij}{\rm LSE}  \{ -\frac{1}{2} ({\bm D_k} - {\bm D_{ij}})^T (h^2 \mathbfss{S}_k +\mathbfss{S}_{ij})^{-1} ({\bm D_k} - {\bm D_{ij}})   \\
 &   +  \ln \frac{f_{\rm B} a_i b_j}{2 \pi \sqrt{\det (h^2 \mathbfss{S}_k +\mathbfss{S}_{ij}) } }  \} .
\end{split} 
\end{align}

\subsection{Priors}

As described in the preceding sections, our general model has 13 parameters,
${\bm \theta} = (\gamma, k, M_0, a_1, a_2, a_3, \dot a_1, \dot a_2, \dot a_3, f_{\rm B}, f_{\rm O}, h_0, h_1)$. To compute
the posterior probability distribution for ${\bm \theta}$, we must define a prior 
distribution, $P({\bm \theta})$. We assume that this distribution is separable
for most parameters, except $a_k$ and $\dot a_k$, which we couple to limit the total
change in shape of $P(q)$ along the main sequence. Additionally, we require $h_1$ to be positive,
and scale its potential distribution with $h_0$.
In the absence of any compelling previous knowledge, and after some trial and error,
we generally adopt sensible uniform, normal or truncated-normal distributions as follows:
\begin{equation}
\label{eqn:prior}
\begin{split}
    P(\log_{10} k) = {} & \mathcal{N}(2,0.3) \\
    P(M_0) = {} & \mathcal{N}(M_{\rm min}+\Delta M/2,\Delta M/2) \\
    P(\gamma) = {} & \mathcal{N}(0,1) \\
    P(a_i) = {} & \mathcal{N}(0,2) \\
    P(\dot a_i) = {} & \mathcal{N}(0,0.1 /\Delta M)) \\
    P(f_B) = {}& U(0.02,0.95) \\
    P(f_O) = {}& U(0,0.05) \\
    P(\log_{10} h_0) = {}&  \mathcal{N}(0.1,0.3) \\
    P(h_1) = {}& \mathcal{N_T}(0,0.4h_0,0,2),
\end{split}
\end{equation}
where $M_{\rm min}$ is the minimum main-sequence mass, $\Delta M$ is half of the mass range of the CMD main sequence, $U(a,b)$ is the uniform distribution between $a$ and $b$, and $\mathcal{N_T}(a,b,c,d)$ is the normal distribution centred at $a$ with standard deviation $b$ truncated below $c$ and above $d$.

\section{Results and conclusions}\label{sec:results}

\begin{figure*}
	\includegraphics[width=2\columnwidth]{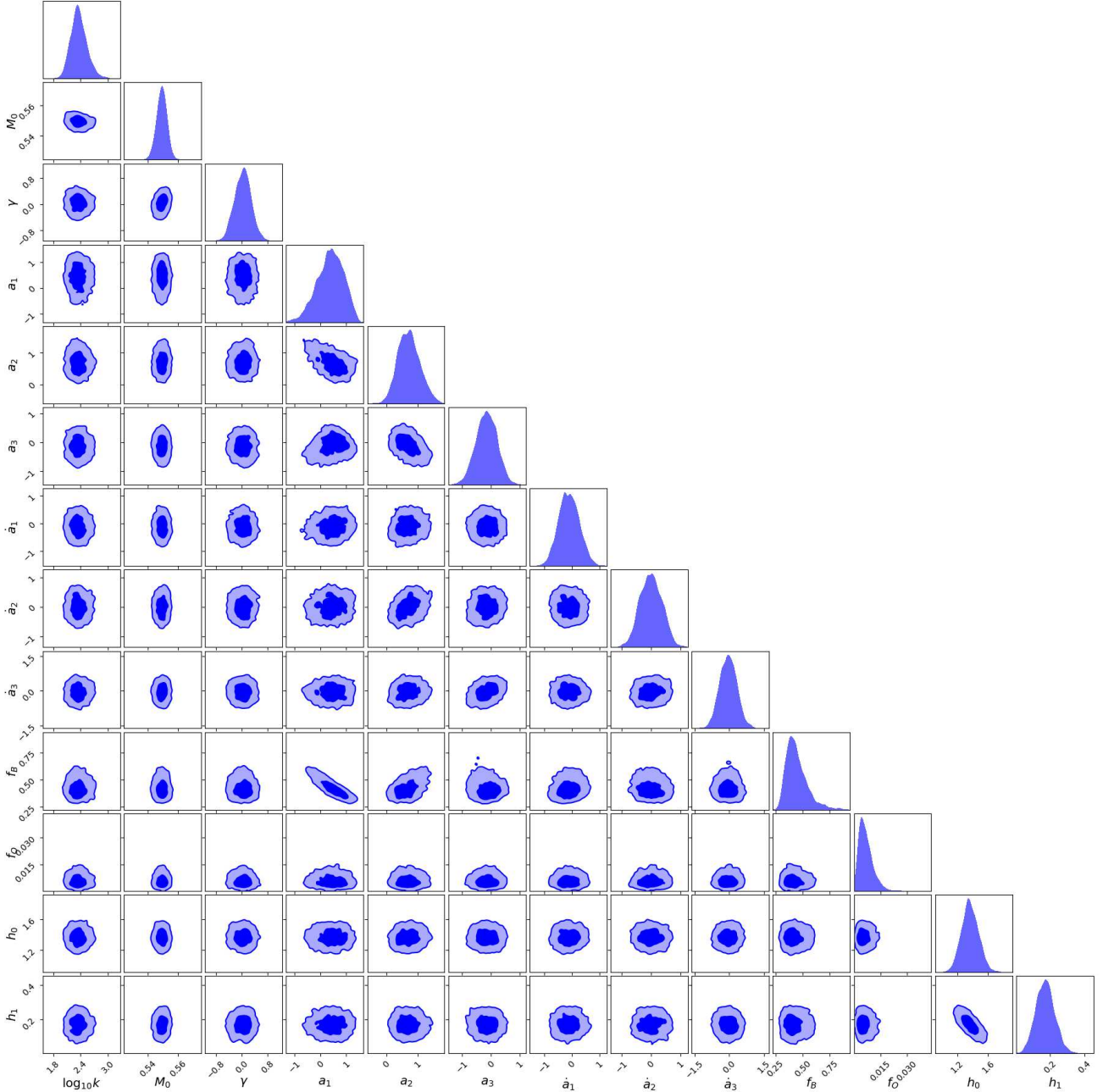}
    \caption{Corner plot for the most general model, model B, showing the 1- and 2-$\sigma$ contours of the posterior probability distribution projected against each pair of parameters (marginalised over the remaining parameters). The 1-D histograms show the marginalised probability distributions for each individual parameter.
    }
    \label{fig:corner}
\end{figure*}

\begin{figure*}
	\includegraphics[width=2\columnwidth]{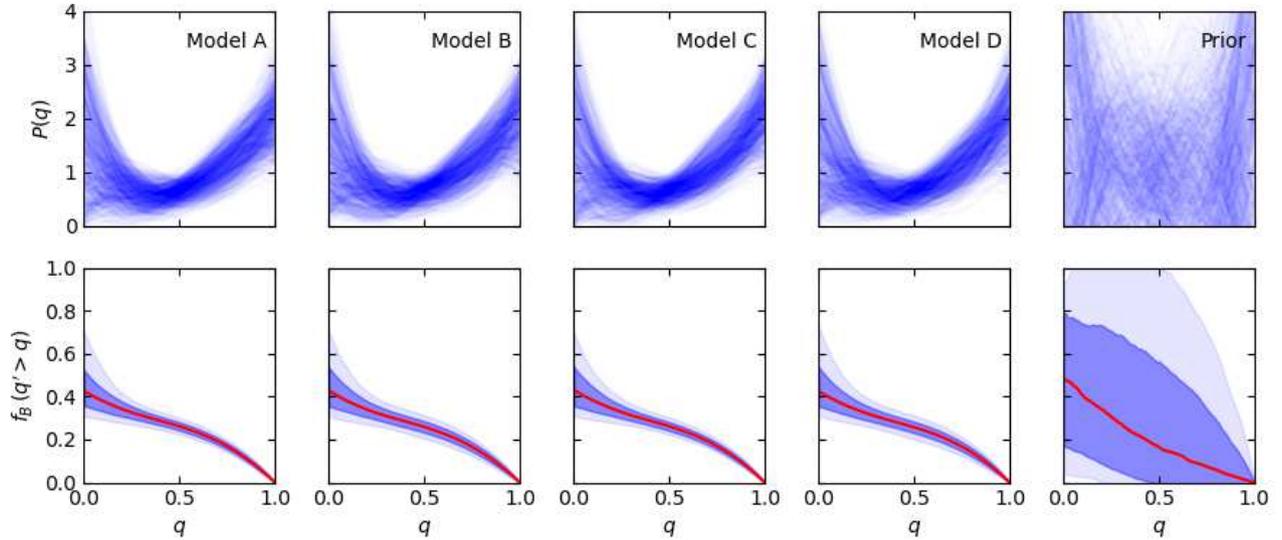}
    \caption{Summary of the mass ratio distribution function and binary fraction for the different models. Each of the first four columns represents a different model, corresponding to the tabulated parameters in Table~\ref{table:results}. 
    The upper row shows the mass ratio distribution
    function for 1000 random weighted samples from the posterior distribution.  The lower row shows the fraction of binary stars with mass ratio greater than $q$ for each model, with the red line representing the median of the posterior distribution and the shaded regions indicating the 1- and 2-sigma
    uncertainties in this quantity.
    The final column shows the same information, but obtained from random sampling of the prior distribution.}
    \label{fig:q_distribution}
\end{figure*}

\begin{figure*}
	\includegraphics{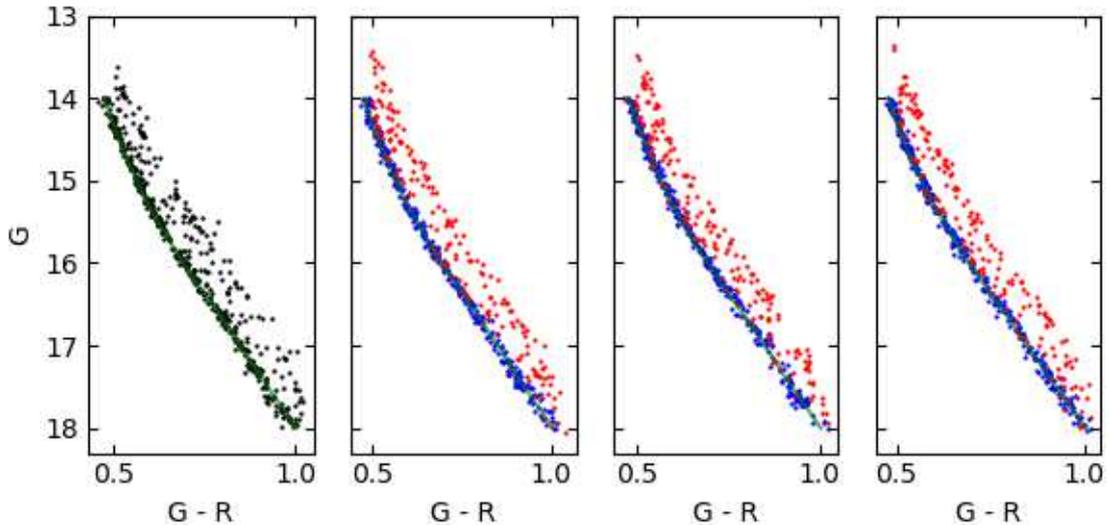}
    \caption{Colour-magnitude diagrams for the data (left panel) and three random realisations of the posterior maximum for model A. Single stars are shown in blue, binaries in red.  
    }
    \label{fig:CMD_realisation}
\end{figure*}

\begin{table} 
	\centering
	\caption{Posterior parameter values and 1-$\sigma$ uncertainties.} \label{table:results}
	\tiny
	\begin{tabular}{lcccc}
\hline
 & A  & B  & C  & D \\
\hline
$\log_{10} k$ & $ 2.37_{-0.16}^{+0.18} $ & $ 2.35_{-0.16}^{+0.18} $ & $ 2.40_{-0.18}^{+0.19} $ & $ 2.37_{-0.16}^{+0.18} $\\
\\
$M_0$ & $ 0.549_{-0.0034}^{+0.0031} $ & $ 0.549_{-0.0034}^{+0.0032} $ & $ 0.5490_{-0.0034}^{+0.0030} $ & $ 0.549_{-0.0032}^{+0.0031} $\\
\\
$\gamma$ & $ 0.04_{-0.25}^{+0.25} $ & $ 0.04_{-0.27}^{+0.24} $ & $ 0.03_{-0.24}^{+0.25} $ & $ 0.03_{-0.24}^{+0.24} $\\
\\
$a_1$ & $ 0.4_{-0.5}^{+0.5} $ & $ 0.4_{-0.6}^{+0.5} $ & $ 0.4_{-0.5}^{+0.5} $ & $ 0.4_{-0.6}^{+0.5} $\\
\\
$a_2$ & $ 0.71_{-0.30}^{+0.34} $ & $ 0.7_{-0.33}^{+0.36} $ & $ 0.71_{-0.30}^{+0.35} $ & $ 0.7_{-0.33}^{+0.39} $\\
\\
$a_3$ & $ -0.1_{-0.4}^{+0.4} $ & $ -0.1_{-0.4}^{+0.4} $ & $ -0.1_{-0.4}^{+0.4} $ & $ -0.1_{-0.4}^{+0.4} $\\
\\
$\dot{a}_1$ & $0$ & $ -0.1_{-0.4}^{+0.4} $ & $0$ & $ -0.1_{-0.4}^{+0.4} $\\
\\
$\dot{a}_2$ & $0$ & $ 0.0_{-0.4}^{+0.4} $ & $0$ & $ 0.0_{-0.4}^{+0.4} $\\
\\
$\dot{a}_3$ & $0$ & $ 0.0_{-0.4}^{+0.4} $ & $0$ & $ 0.0_{-0.4}^{+0.4} $\\
\\
$f_B$ & $ 0.43_{-0.07}^{+0.10} $ & $ 0.43_{-0.07}^{+0.11} $ & $ 0.43_{-0.07}^{+0.11} $ & $ 0.42_{-0.07}^{+0.11} $\\
\\
$f_O$ & $ 0.0060_{-0.0028}^{+0.0044} $ & $ 0.0061_{-0.0029}^{+0.0043} $ & $ 0.0062_{-0.0030}^{+0.0045} $ & $ 0.0063_{-0.0030}^{+0.0044} $\\
\\
$h_0$ & $ 1.36_{-0.10}^{+0.11} $ & $ 1.36_{-0.10}^{+0.11} $ & $ 1.69_{-0.08}^{+0.08} $ & $ 1.69_{-0.07}^{+0.08} $\\
\\
$h_1$ & $ 0.17_{-0.05}^{+0.05} $ & $ 0.17_{-0.05}^{+0.05} $ & $0$ & $0$\\
\\
\hline
$\Delta \ln P$ & $ 0.00 $ & $ -0.01 $ & $ -4.49 $ & $ -4.49 $\\
$\Delta \log_{10} Z$ & $ 0.00 $ & $ -0.04 $ & $ -3.08 $ & $ -3.25 $\\
\hline
\end{tabular}

\end{table}

\begin{table} 
	\centering
	\caption{Fraction of binary stars with mass ratio greater than a given $q$ for each model.} \label{table:f_tab}
	\tiny      
	\begin{tabular}{lcccc}
\hline
$q$ & A  & B  & C  & D \\
\hline
0.1 & $ 0.37_{-0.05}^{+0.05} $ & $ 0.37_{-0.05}^{+0.06} $ & $ 0.38_{-0.05}^{+0.06} $ & $ 0.37_{-0.05}^{+0.06} $\\
\\
0.2 & $ 0.34_{-0.033}^{+0.034} $ & $ 0.34_{-0.034}^{+0.039} $ & $ 0.34_{-0.033}^{+0.037} $ & $ 0.33_{-0.04}^{+0.04} $\\
\\
0.3 & $ 0.310_{-0.024}^{+0.024} $ & $ 0.307_{-0.029}^{+0.030} $ & $ 0.310_{-0.025}^{+0.027} $ & $ 0.305_{-0.029}^{+0.031} $\\
\\
0.4 & $ 0.284_{-0.020}^{+0.020} $ & $ 0.282_{-0.027}^{+0.027} $ & $ 0.285_{-0.022}^{+0.021} $ & $ 0.280_{-0.026}^{+0.027} $\\
\\
0.5 & $ 0.258_{-0.019}^{+0.019} $ & $ 0.256_{-0.026}^{+0.026} $ & $ 0.258_{-0.020}^{+0.020} $ & $ 0.254_{-0.026}^{+0.026} $\\
\\
0.6 & $ 0.227_{-0.018}^{+0.018} $ & $ 0.224_{-0.025}^{+0.025} $ & $ 0.228_{-0.019}^{+0.019} $ & $ 0.224_{-0.026}^{+0.025} $\\
\\
0.7 & $ 0.188_{-0.016}^{+0.016} $ & $ 0.185_{-0.022}^{+0.023} $ & $ 0.188_{-0.017}^{+0.017} $ & $ 0.186_{-0.023}^{+0.023} $\\
\\
0.8 & $ 0.136_{-0.013}^{+0.014} $ & $ 0.134_{-0.019}^{+0.019} $ & $ 0.137_{-0.014}^{+0.015} $ & $ 0.136_{-0.019}^{+0.019} $\\
\\
0.9 & $ 0.071_{-0.009}^{+0.010} $ & $ 0.070_{-0.012}^{+0.013} $ & $ 0.072_{-0.010}^{+0.010} $ & $ 0.071_{-0.012}^{+0.013} $\\
\hline
\end{tabular}

\end{table}

We have investigated four different models that use or do not use $\dot a_k$
and $h_1$ in various combinations.

To sample the posterior distributions (Eqn.~\ref{eqn:final_ln_prob}) we have used the affine-invariant
ensemble sampler EMCEE \citep{Foreman-Mackey2013}, and the 
nested sampler \citep{Skilling2004} DYNESTY \citep{Higson2019,Speagle2020}. Fundamentally these are different approaches to sampling the probability space. EMCEE, a Markov Chain Monte Carlo sampler, starts with a number of points (walkers) distributed close to a guess at the posterier maximum probability. It then uses a quasi-downhill iteration on $-\ln P$ to roughly locate the maximum. The walkers converge to a stationary distribution that mirrors the probability distribution. In contrast, the nested sampler begins with points distributed throughout the entire prior hypervolume, iteratively replacing points that are outside a rising probability contour. Despite the different methods, both samplers converged to the same results.
Following this process, the Nelder-Mead method was used to locate the maximum posterior probability point for each model.

In Table~\ref{table:results} we list the parameter medians and one-sigma uncertainties for our four models that include or exclude the  combinations of $\dot a_k$ and $h_1$. 
Posterior parameter distributions for all parameters for the most general model (model B) are shown as marginalised corner plots and histograms in Fig.~\ref{fig:corner}. The parameters are well defined but, as might be expected, there are covariances between $f_B$ and the shape parameters for the $q$ distribution (i.e. we would expect $q$ distributions that rise towards $q=0$ to result in a higher value of $f_B$).
Corner plots for the more restricted models are similar.

Also listed in Table~\ref{table:results} are two extra values for each model. Firstly, we quote the 
the relative maximum probability, $\Delta \ln P({\bm \theta} | {\bm D})$, for each model. This is a measure of how well the model fits the data. For readers more familiar with the $\chi^2$ statistic, $\Delta \ln P$ is equivalent to $\Delta \chi^2/2$.
Secondly, we quote the Bayes factor, $\Delta \log_{10} Z$, which indicates the overall relative probability of the
different models given the data, compensating implicitly for the extra fit freedom that is introduced with additional parameters. An interpretation of $\Delta \log_{10} Z$ is given by \citet{Kass1995}, that $\Delta \log_{10} Z > (0.5, 1, 2)$ represents ("substantial", "strong", "decisive") evidence for a proposition. The quoted $\Delta \log_{10} Z$
values have an estimated uncertainty of 0.2.

In Fig.~\ref{fig:q_distribution} we summarize the implied results for the mass ratios and binary fractions. For 
each model we use random samples from the posterior parameter distribution to show the mass ratio distribution function and the fraction of binary stars with a mass ratio greater than 
a given $q$, 
\begin{equation}
f_B(q' > q) = f_B \int_q^1 P(q|a_k,\dot{a}_k,M; k = 1..3) \, dq.
\end{equation}
This latter quantity is tabulated in Table~\ref{table:f_tab} for equally spaced values of $q$.
Readers may use this table to find binary fractions with mass ratios above any particular desired limit.
The distributions implied by the prior, Equations~\ref{eqn:prior}, are shown for comparison, demonstrating that it is the likelihood, not the prior, that is driving the results.

In Fig.~\ref{fig:CMD_realisation} we show our final data, along with three random realisations of model A with the posterior maximum probability parameters.

From these results we conclude the following:

\begin{enumerate}
    \item
    There is decisive evidence that models A and B, with error bar scaling that increases towards the lower main sequence ($h_1$), are better than those with constant scaling. We remind the reader that an  error bar scaling greater than 1 is equivalent to allowing an intrinsic width to the main sequence, so that models A and B can also be interpreted as allowing for an
    increasing main-sequence width towards fainter magnitudes.
    \item
    All of the models have 
    a rising mass-ratio distribution function with $q$  for $q > 0.5$ (Fig.~\ref{fig:q_distribution}). This shows that more binary stars (per unit $q$) 
    exist with high mass ratios than intermediate mass ratios, but not with the extreme peak found by \citet{Fisher2005}.
    \item
    Below $q \approx 0.3$, there is little constraint on the form of the mass ratio distribution function. This is to be expected, since such binary stars lie on or very close to the main sequence and cannot be distinguished from single stars.
    \item
    The implied binary fraction is well constrained for $q \gtrsim 0.3$, but becomes increasingly less so
    as $q \rightarrow 0$.
    \item
    The "best-fitting" model is model A, which has a constant mass ratio distribution function shape. However, the improvement of fit over the other model (B) that includes $h_1$ is negligible ($\Delta \ln P = 0.01$).
    \item
    The model with the strongest evidence, $Z$, is also model A with a constant shape to the $q$ distribution.
    However, the evidence that this model is better than model B (which has a mass-varying shape) is insubstantial. Having fewer parameters, we adopt model A as our favoured option. 
    \item
    Since the fraction of outliers, $f_O$, is negligible for each model, we can interpret $f_B$ as being the binary
    frequency, without need for adjustment.
    \item
    Formally, the favoured models A and B imply an overall binary fraction, $f_B = 0.43_{-0.07}^{+0.11}$, but this 
    result may be unreliable. The lower limit is sensible, and is imposed by the reliable detection of 
    binaries with $q > 0.5$, coupled with the realistic expectation that the binary frequency below this threshold is not zero. However the reason for the upper limit is not obvious, yet is nonetheless required by the data. Larger 
    values of $f_B$ could eventuate if the binary mass ratio distribution function were to rise more steeply towards $q=0$ than is apparent in Fig.~\ref{fig:q_distribution}. Such behaviour is permitted by the prior (see the final column of Fig.~\ref{fig:q_distribution}), yet is not favoured by the (data-driven) likelihood. We are unable to pinpoint a particular physical cause.
    Restricting ourselves to higher-mass-ratio binaries, we find $f_B(q > 0.5) = 0.258 \pm 0.019$.
\end{enumerate}

\section{Summary}

We have presented a new generative model for star-cluster colour magnitude diagrams that allows the measurement of the binary star mass ratio distribution as well as the binary star frequency. The model naturally accounts for the locations of stars on the CMD without individual classification of stars as single or binaries.

The first application of the model is presented using Gaia photometry of the old open cluster, M67.
By considering a 5-dimensional phase-space distribution of stars in the direction of M67, we have obtained a very clean sample of cluster stars that cover the low-main sequence through to just above the main-sequence turn-off.

We find the frequency of binary stars in M67 with mass ratios greater than 0.5 to be $f_B(q > 0.5) = 0.258 \pm 0.019$. The mass ratio distribution is found to rise gently towards higher mass ratios.
There is no compelling evidence that the form of the distribution varies with primary mass along the main sequence.

\section*{Acknowledgements}

We are grateful for the comments of an anonymous referee, who prompted us to consider the $q$ detection threshold.

\section*{Data and Code Availability}

Gaia Early Data Release 3 (EDR3) data is publicly available via the Gaia archive,  https://gea.esac.esa.int/archive/ ,
and the Centre de Donn\'ees astronomiques de Strasbourg(CDS) catalogue service, https://vizier.cds.unistra.fr/viz-bin/VizieR.

The code used for this analysis is written in PYTHON and CUDA (via the pyCUDA python library). 
CUDA is an extension to C/C++ that uses an NVIDIA graphical processing unit to perform
parallel calculations. We use CUDA to perform the likelihood calculation from Section~\ref{sec:total_likelihood}. The CUDA language exposes GPU hardware cores
as groups of threads arranged into blocks.
In our code, GPU threads are used to sum over basis functions, and blocks 
to sum over data points. The code 
is being developed publicly at \url{https://github.com/MichaelDAlbrow/CMDFitter}.



\bibliographystyle{mnras}
\bibliography{references} 




\appendix

\section{CMD covariance}
\label{appendixA}

The covariance, $\sigma_{G-R,G}$ can be calculated from the definition of covariance and the algebra of expectation values ($E$),
\begin{equation}
\begin{split}
\sigma_{G-R,G} = {} & {\rm Cov}(G-R,G) \\
= & E[(G-\mu_G)((G-R) - \mu_{G-R}] \\
  & - E[G-\mu_G] E[ (G-R) - \mu_{G-R} ] \\
= & E[(G-\mu_G)\left( (G-\mu_G) - (R-\mu_R) \right) ] - 0 \times 0 \\
= & E[(G-\mu_G)^2 - (G-\mu_G)(R-\mu_R) ] \\
= & E[(G-\mu_G)^2] - E[(G-\mu_G)(R-\mu_R) ] \\
= & E[(G-\mu_G)^2] - E[(G-\mu_G)] E[(R-\mu_R) ] \\
= & \sigma_G^2 - 0 \times 0 \\
= &  \sigma_G^2 
\end{split}
\end{equation}


\bsp	
\label{lastpage}
\end{document}